\documentclass[pra,superscriptaddress,twocolumn,amsmath,amssymb]{revtex4}
\usepackage{amsfonts}
\usepackage{amssymb}
\usepackage{mathtools}
\usepackage{graphicx}
\usepackage{subfigure}
\usepackage{mathrsfs}
\usepackage{color}
\usepackage[normalem]{ulem}
\usepackage{natbib}
\usepackage{bbold}
\usepackage{placeins}
\usepackage{braket}
\usepackage{soul,xcolor}
\usepackage{epstopdf}
\usepackage{tabu}
\usepackage{array}
\usepackage{bm}
\usepackage{upgreek}
\usepackage{multirow}
\usepackage[utf8]{inputenc}
\usepackage[colorlinks = truelinkcolor = blue, urlcolor  = blue, citecolor = blue, anchorcolor = blue]{hyperref}
\hypersetup{
	colorlinks   = true, 
	urlcolor     = blue, 
	linkcolor    = blue, 
	citecolor   = blue 
}

\begin{document}
\title{A study of coherence based measure of  quantumness in (non) Markovian channels}

	\author{Javid Naikoo}
	\email{naikoo.1@iitj.ac.in}
	\affiliation{Indian Institute of Technology Jodhpur, Jodhpur 342011, India}

	\author{Subhashish Banerjee}
	\email{subhashish@iitj.ac.in}
	\affiliation{Indian Institute of Technology Jodhpur, Jodhpur 342011, India}

	
	\begin{abstract}
     We make a detailed analysis  of  quantumness  for various quantum noise channels, both Markovian and non-Markovian. The noise channels considered include dephasing channels like random telegraph noise, non-Markovian dephasing and phase damping, as well as the non-dephasing channels such as generalized amplitude damping and  Unruh channels.  We make use of a recently introduced  witness for quantumness  based on the square $l_1$ norm of coherence. It is found that the increase in the degree of non-Markovianity increases the quantumness of the channel.
 	\end{abstract}

	\maketitle 
	
	\section{Introduction}
Quantum coherence \cite{hu2018quantum,wilde2013quantum} is central to quantum mechanics, playing a fundamental role for  the manifestations of quantum properties of a system.  It is at the heart of the phenomena such as multi-particle interference and entanglement which are  pivotal for carrying out various quantum information and  communication tasks, viz., quantum key distribution \cite{bennett1992experimental,grosshans2003quantum} and teleportation \cite{bennett93teleport}. An operational formulation of coherence as a resource theory was recently developed \cite{Plenio201resource}. The notion of coherence \cite{Bryne2018} has its roots in quantum optics \cite{Glauber1963,Sudarshan1963}. Recent developments have made use of coherence in superconducting systems \cite{almeida2013probing},  biological systems \cite{lloyd2011quantum}, non-Markovian phenomena \cite{bhattacharya2018evolution}, foundational issues \cite{cohmix,yadin2016general}  and subatomic physics \cite{alok2016quantum,dixit2019study}. 

Quantum channels are completely positive (CP) and trace preserving maps between the spaces of operators which can transmit classical as well as quantum information.
Quantum information protocols are based on the fact that information is transmitted in the  form of quantum states. This is achieved either by directly sending non-orthogonal states or by using pre-shared entanglement. The channels can reduce the degree of coherence and entanglement as the information flows from sender to receiver. Interestingly, it was shown in \cite{Karimpour2015} that quantum channels can have  cohering power and that  a qubit unitary map has equal cohering and decohering power in any basis. In general, the extent to which the quantum features are affected depends on the underlying dynamics and the type of noise.  Therefore it is natural to ask to what extent is  coherence preserved by a channel used to transmit  quantum information. 

 The physical foundation of a large number of  quantum channels relies on the Born and/or Markov approximations \cite{Maniscalco}. However, in a number of quantum communication tasks, the characteristic time scales of the system of interest become comparable with the reservoir correlation time. Therefore, a non-Markovian description for such scenarios becomes indispensable \cite{banerjee2018open}.
 
 A quantum channel can be used to transport either classical or quantum information. The reliability of a quantum channel is tested by the probability that the output and input states are the  same. A well known measure to quantify the performance of a channel is the \textit{average fidelity} \cite{braunstein2000criteria,horodecki1995violating,horodecki1996teleportation, horodecki1999SingletFraction}.  The notion of \textit{fidelity} of two quantum states  provides a qualitative measure of their distinguishability \cite{jozsa1994fidelity}.
 
  In \cite{Shahbeigi2018}, a  witness of nonclassicality of a channel was introduced. This  is  based on average quantum coherence of the state space, using the square $l_1$ norm of coherence of qubit channels. It was shown that the extent to which quantum correlation is preserved under local action of the channel cannot exceed the quantumness of the underlying channel.

In this work, we use the definition of quantumness based on the average coherence and apply it to different channels, both Markovian and non-Markovian. Average channel fidelity is a useful figure of merit when considering channel transmission, particularly in the presence of noise. Accordingly, a corresponding  study is made  on these channels. The paper is organized as follows. In Sec. (\ref{secII}), we briefly review the definition of nonclasscality of quantum channels. Section (\ref{channels}) is devoted to analyzing the interplay of quantumness and average fidelity in various noise models. Results and their discussion is made in Sec (\ref{resultdiscussion}). We conclude in Sec. (\ref{conclusion}).

\section{Quantum channels and the measure of quantumness}\label{secII}
Here, we given a brief overview of quantum channels. This will be followed by a discussion on a coherence based measure of quantumnes of channels. 
\subsection{Quantum channel}
A quantum channel in the Sch\"odinger picture is a completely positive and trace preserving map $\Phi : \mathcal{T} (\mathcal{H}_A) \rightarrow \mathcal{T} (\mathcal{H}_B)  $, where $\mathcal{H}_A$ and $\mathcal{H}_B$ are the underlying Hilbert spaces for system $A$ and $B$, respectively. One defines a \textit{dual} channel, in the Heisenberg picture, as a linear, completely positive map $\Phi^* = \mathcal{S} (\mathcal{H}_A) \rightarrow \mathcal{S} (\mathcal{H}_B)$. When the input and output systems have equal dimensions ($d_A = d_B$), $\Phi^*$ is also trace-preserving, so that both $\Phi$ and $\Phi^*$ are channels in the Sch\"{o}dinger and Heisenberg pictures \cite{holevo2012quantum}.\par
The operator sum representation of a channel is given as 
\begin{equation}\label{map}
\Phi [\bm{\rho}]= \sum\limits_{\mu} \textbf{M}_\mu \bm{\rho} \textbf{M}^{\dagger}_{\mu},
\end{equation}
such that the operators $\textbf{M}_\mu$, called as \textit{Kraus operators}, obey the completeness condition, $\sum\limits_{\mu} \textbf{M}_{\mu}^\dagger \textbf{M}_{\mu} = \mathbb{1}$. Here, $\mathbb{1}$ is the identity operator. Note that $\bm{\rho}$, in Eq. (\ref{map}), need not be a pure state. A linear map given in Eq. (\ref{map}), is called a \textit{quantum channel} or \textit{superoperator} (as it maps operators to operators) or \textit{completely positive trace preserving} (CPTP) map. A quantum channel is characterized by the following properties (i) \textit{linearity}, i.e., $\Phi [\alpha \bm{\rho}_1 + \beta \bm{\rho}_2] = \alpha \Phi (\bm{\rho}_1) + \beta \Phi (\bm{\rho}_2)$, where $\alpha$ and $\beta$ are complex number, (ii)  Hermiticity preserving, i.e., $\bm{\rho} = \bm{\rho}^\dagger \implies \Phi[\bm{\rho} ] = \Phi[\bm{\rho}]^\dagger$, (iii)  positivity preserving, i.e., $\bm{\rho} \ge 0 \implies \Phi[\bm{\rho}] \ge 0$, and (iv)  trace preserving, i.e., $Tr(\Phi[\bm{\rho}]) = Tr(\bm{\rho})$. 

\subsection{A coherence based measure of quantumness of channels}\label{quantumness}
A coherence based measure was introduced in \cite{Shahbeigi2018}
\begin{equation}\label{generalQ}
Q_C(\Phi) = N_C \int C(\Phi(\rho)) d \mu(\rho).
\end{equation}
Here, $\Phi$ is the channel under consideration, $C$ denotes the chosen measure of coherence and $N_C$ is a normalization constant. To proceed, we analyze the effect of a qubit channel on  a state \( \bm{\rho} = \frac{1}{2} (\bm{I} + \bm{\xi } \bm{\sigma} ) \). This turns out to be a \textit{affine} transformation on a Bloch sphere, such that the Bloch vector $\xi$ transforms as
\begin{equation}
\Phi(\bm{\rho})  = \bm{\rho}^\prime =  \frac{1}{2} (\bm{I} + \bm{\xi}^\prime \bm{\sigma} ).
\end{equation}
Here, $\bm{\xi}^\prime = \bm{A} \bm{\xi} + \bm{B}$,	such that the matrices $\textbf{A}_{3 \times 3}$ and $\textbf{B}_{3 \times 1}$ depend on the channel parameters. By choosing the square $l_1$-norm as the measure of coherence, we compute the coherence with respect to an arbitrary orthonormal basis. To make the quantumness witness a basis independent quantity, one performs  optimization over all orthonormal basis, leading to  a closed expression for the quantumness witness
\begin{equation}\label{Qdef}
Q_{C_{l_1}^{2}} (\Phi) = \lambda_2 + \lambda_3.
\end{equation}  
Here, $\lambda_1 \ge \lambda_2 \ge \lambda_3$ are eigenvalues of matrix  $\bm{\mathcal{L}} = \frac{1}{2} (\bm{A} \bm{A}^T + 5 \bm{B} \bm{B}^T)$, with $T$ denoting the transpose operation. Thus, Eq. (\ref{Qdef}) gives an operational definition of the quantumness of a channel. In what follows, we will drop the subscript $C_{l_1}^2$ and call the quantumness of a map $\Phi$ just as $Q(\Phi)$. It is worth  mentioning here, that for the unital channels, which map identity to identity, i.e., $\Phi(\bm{I}) = \bm{I}$, the above definition of quantumness coincides with the geometric discord \cite{Shahbeigi2018}.

\section{Specific channels}\label{channels}
In this section, we give a brief account of various quantum channels \cite{nielsen2002quantum,watrous2018theory} used in this work. The dephasing class includes random telegraph noise (RTN) \cite{Daffer,PradeepOSID,Pradeep2}, non-Markovian dephasing (NMD) \cite{shrikant2018non} and phase damping (PD) \cite{banerjee2007dynamics} channels while  in the non-dephasing class, we consider generalized amplitude damping (GAD) \cite{srikanth2008squeezed,omkar2013dissipative} and Unruh channels \cite{UnruhSB}. \par
\textit{Random Telegraph Noise}: This channel characterizes the  dynamics when the system is subjected to a bi-fluctuating classical noise, generating RTN with pure dephasing.  The dynamical map acts as follows
\begin{equation}
 \Phi^{RTN} (\bm{\rho}) = \bm{\mathcal{R}}_0 \bm{\rho} \bm{\mathcal{R}}_0^\dagger + \bm{\mathcal{R}}_1 \bm{\rho} \bm{\mathcal{R}}_1^\dagger,
 \end{equation}
  where the two  Kraus operators are given by
\begin{equation}
\bm{\mathcal{R}}_0 = \sqrt{\frac{1 + \Lambda(t)}{2}} \bm{I}, \qquad \bm{\mathcal{R}}_1 = \sqrt{\frac{1 - \Lambda(t)}{2}} \bm{\sigma}_z.
\end{equation}
Here, $\Lambda(t)$ is the \textit{memory~kenel}
\small
\begin{equation}\label{Lambdadef}
\Lambda (t) = e^{-\gamma t} \Bigg[\cos\Big[ \sqrt{\big(\frac{2 b}{\gamma} \big)^2 -1}~ \gamma  t \Big] + \frac{\sin \Big[ \sqrt{ \big(\frac{2 b}{\gamma} \big)^2 -1}~ \gamma  t \Big]}{\sqrt{\big(\frac{2 b}{\gamma} \big)^2 -1}} \Bigg],
\end{equation}
\normalsize
 where $b$ quantifies the system-environment coupling strength and $\gamma$ is proportional to the fluctuation rate of the RTN. Also, $\bm{I}$ and $\bm{\sigma}_z$ are the identity and Pauli spin matrices, respectively. The completeness condition reads $\bm{\mathcal{R}}_0 \bm{\mathcal{R}}_0^\dagger + \bm{\mathcal{R}}_1 \bm{\mathcal{R}}_1^\dagger = \bm{I}$. The dynamics is Markovian [non-Markovian] if $(4b \uptau)^2 > 1 ~[ (4b \uptau)^2 < 1 ]$, where $\uptau = 1/(2\gamma)$. Starting with the state $ \bm{\rho} = \frac{1}{2} (\bm{I} + \bm{\xi } \bm{\sigma} )  $, the new Bloch vector is given by $\bm{\xi}^\prime = [ \bm{\xi}_x \Lambda(t), \bm{\xi}_y \Lambda(t), \bm{\xi}_z ]^T$. This implies $\bm{A} = {\rm diag.}[\Lambda(t), \Lambda(t),1]$ and $\bm{B} = 0$, and consequently, $\bm{\mathcal{L}} = {\rm diag.} [\frac{1}{2}[\Lambda (t)]^2, \frac{1}{2} [\Lambda (t)]^2, \frac{1}{2} ]$. Since, $-1\le \Lambda(t) \le 1$, we identify both the small eigenvalues as $\frac{1}{2} [\Lambda (t)]^2$, leading to 
\begin{equation}\label{QRTN}
Q(\Phi^{RTN}) =  [\Lambda (t)]^2.
\end{equation}
We, next compute the fidelity for the states $\bm{\rho}$ and $\bm{\rho}^\prime$ and study its interplay with  quantumness. The fidelity between  qubit states  $\bm{\rho}$ and $\bm{\rho}^\prime$ is  
\begin{equation}\label{def:fidelity}
F(\bm{\rho}, \bm{\rho}^\prime) = Tr[\bm{\rho} \bm{\rho}^\prime] + 2 \sqrt{Det[\bm{\rho}] Det[\bm{\rho}^\prime]}.
\end{equation}
Using a general qubit parametrization
\begin{equation}
\bm{\rho} = \begin{bmatrix}
                      \cos^2(\theta/2)      & \frac{1}{2}e^{-i \phi} \sin( \theta) \\
                      \frac{1}{2}e^{i \phi} \sin( \theta)   &  \sin^2(\theta/2)
                  \end{bmatrix},
\end{equation}
the fidelity for RTN model turns out to be
\begin{equation}
F_{RTN} = \frac{1}{4} \big[ 3 + \cos(2 \theta) + 2 \sin^2(\theta) \Lambda (t)\big].
\end{equation}
In order to make this quantity state independent, we calculate the \textit{average} fidelity $\mathcal{F} = \frac{1}{4 \pi} \int_{0}^{2 \pi} \int_{0}^{\pi} F \sin(\theta) d\theta d\phi$. We have
\begin{equation}\label{avgFidRTN}
\mathcal{F}_{RTN} = \frac{1}{3} [2 + \Lambda(t)].
\end{equation}
Since $-1 \le \Lambda (t) \le 1$, the average fidelity is symmetric about its classical value $2/3$.
\begin{figure}
	\centering
	\includegraphics[width=60mm]{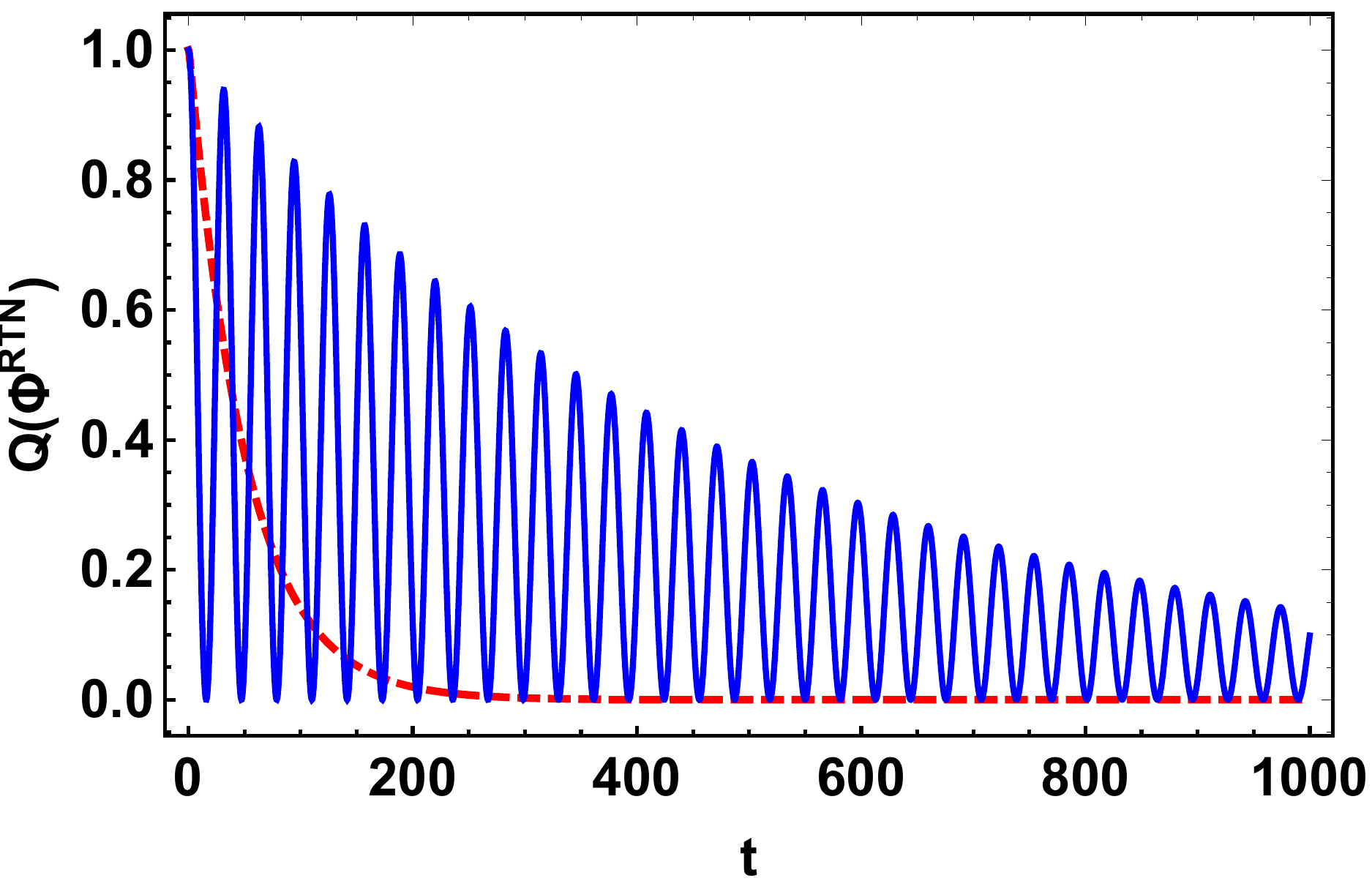}\\
	\includegraphics[width=60mm]{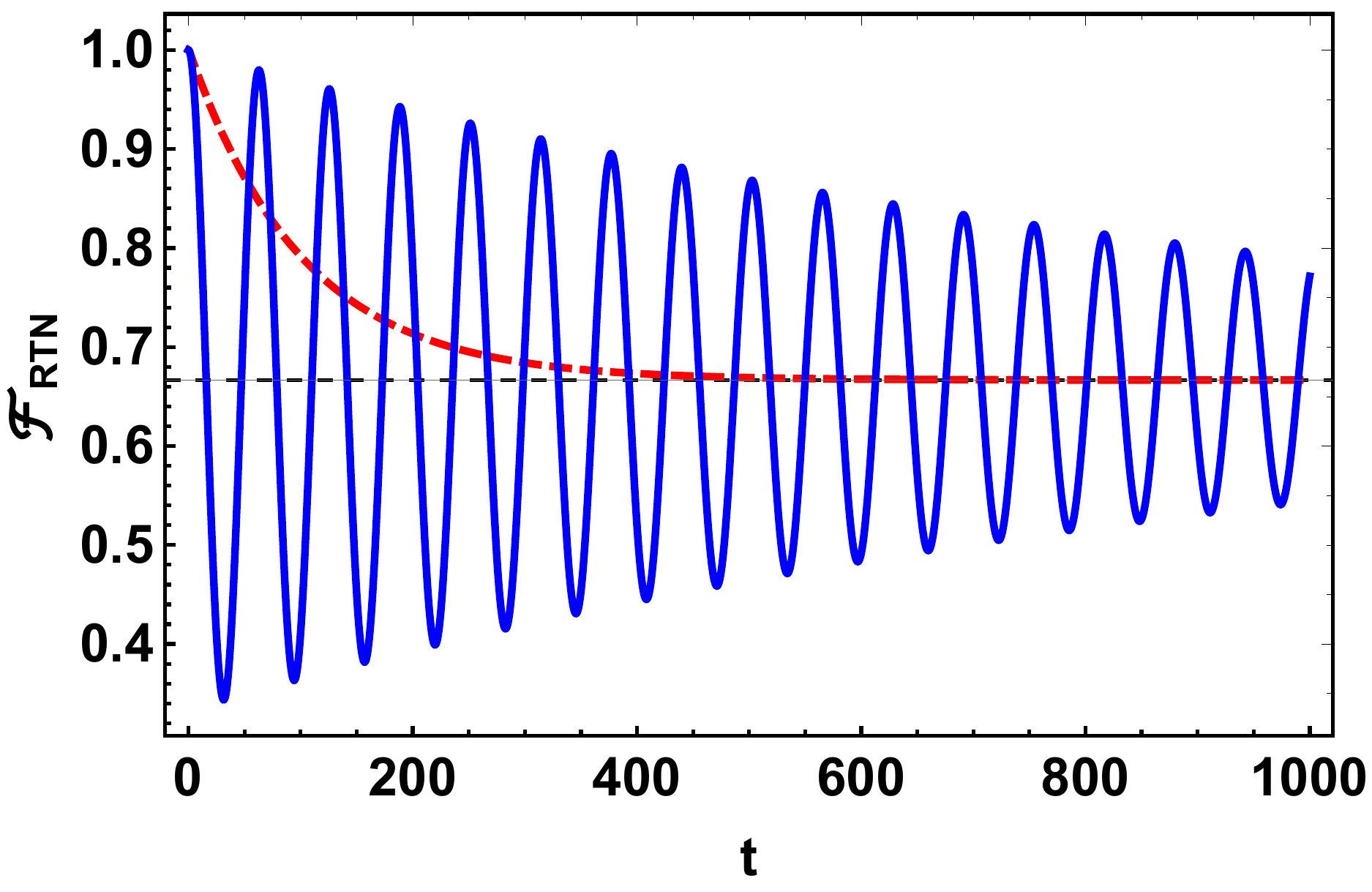}
	\caption{RTN channel: The quantumness $Q(\Phi^{RTN})$ Eq. (\ref{QRTN}) and \textit{average} fidelity $\mathcal{F}_{RTN}$ Eq. (\ref{avgFidRTN}), plotted with respect to time $t$ (sec.), for a qubit subjected to RTN. The solid (blue) and dashed (red) curves correspond to non-Markovian ($b=0.05,~\gamma=0.001$) and Markovian ($b=0.07,~\gamma$) cases, respectively. The fidelity oscillates symmetrically about $2/3$ in non-Markovian case, while in Markovian case, it decreases monotonically and saturates to this value.}
	\label{fig:RTN}
\end{figure}
\FloatBarrier

\textit{Non-Markovian dephasing}: This channel is an extension of the dephasing channel to non-Markovian class. The non-Markovianity is identified with the breakdown in complete-positivity of the map. The Kraus operators are given by
\begin{align}
\bm{\mathcal{N}}_0 &= \sqrt{(1-\alpha p)(1-p)}~ \bm{I}, \nonumber\\ {\rm and} ~~~
 \bm{\mathcal{N}}_1 &= \sqrt{p+\alpha p(1-p)}~ \bm{\sigma}_z.
\end{align}
Here, the parameter $\alpha$ quantifies the degree of non-Markovianity of the channel and $p$  is a time-like parameter such that $0 \le p \le 1/2$.
In this case, the quantumness parameter turns out to be
\begin{equation}\label{QNMD}
Q(\Phi^{NMD}) =  \Omega^2(p),
\end{equation}
where $\Omega = 1 - 2p - 2\alpha p (1-p)$. The average fidelity, in this case, is given by
\begin{equation}\label{avgFidNMD}
\mathcal{F}_{NMD} = \frac{1}{3} [2 + \Omega(p)].
\end{equation}
We use the parametrization $p = \frac{1}{2}(1-e^{-\kappa t})$, such that as $t: 0 \rightarrow \infty$, $p: 0 \rightarrow 1/2$.
\begin{figure}
	\centering
	\includegraphics[width=60mm]{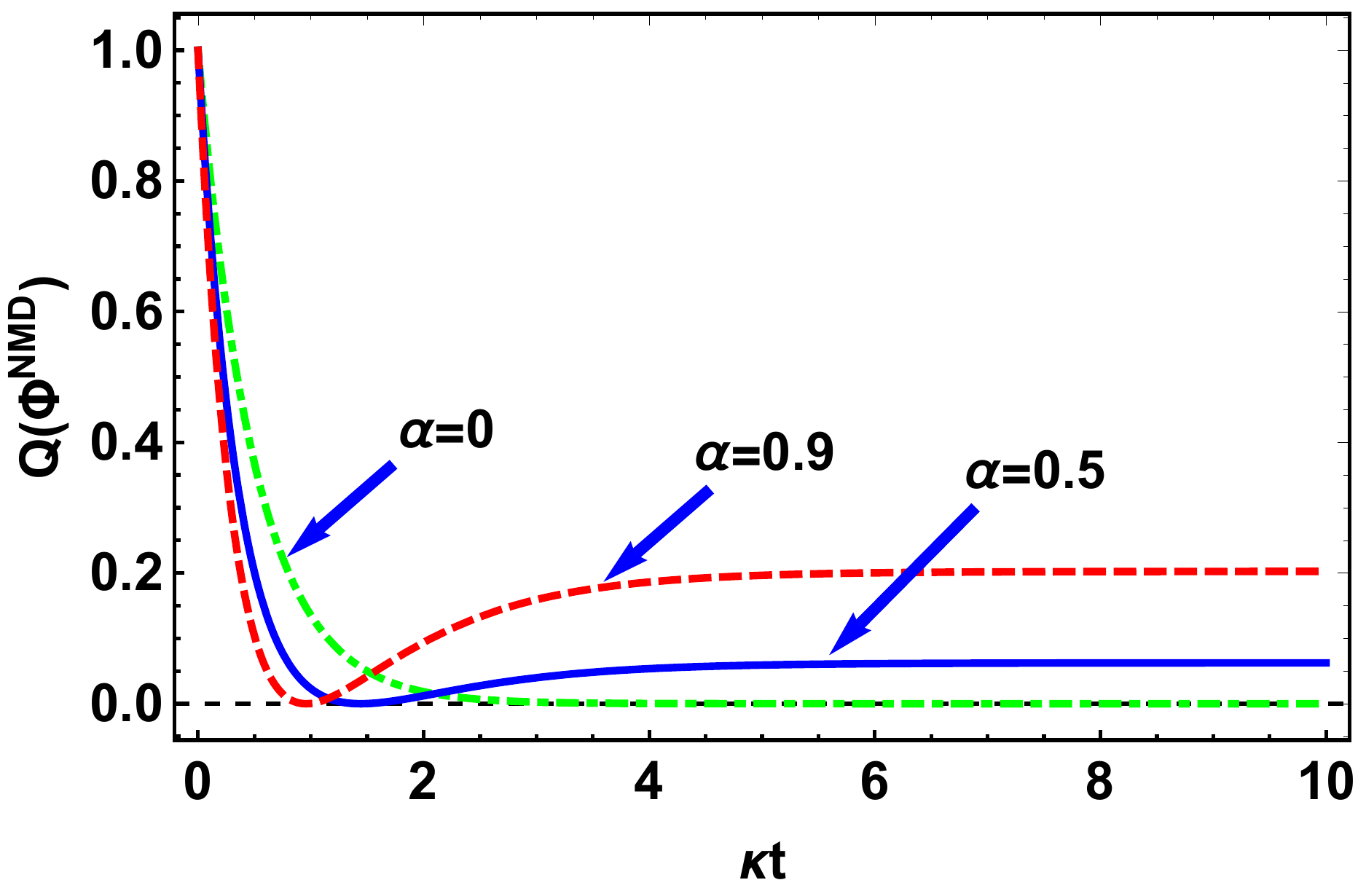}\\
	\includegraphics[width=60mm]{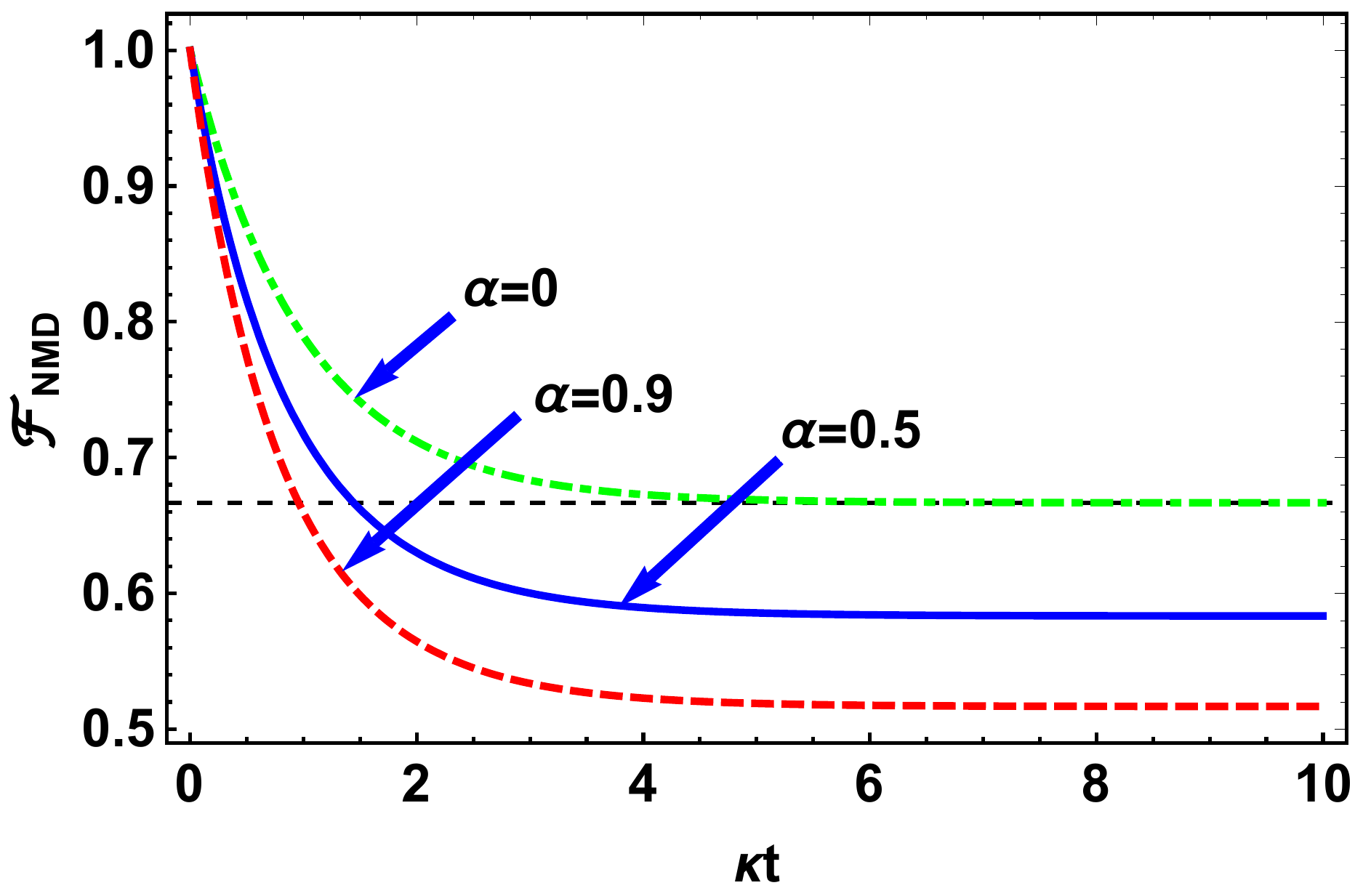}
	\caption{NMD channel: The quantumness $Q(\Phi^{NMD})$ Eq. (\ref{QNMD}) and \textit{average} fidelity $\mathcal{F}_{NMD}$ Eq. (\ref{avgFidNMD}), plotted with respect to the dimensionless  parameter $\kappa t$, for a qubit subjected to NMD, for different values of parameter $\alpha$.}
	\label{fig:NMD}
\end{figure}

\textit{Phase damping (PD) channel}: PD channel models the phenomena where decoherence occurs without dissipation (loss of energy). The dynamical map, in this case has the Kraus representation
\begin{equation}
\bm{\mathcal{P}}_0 = \begin{bmatrix}
                                  1 & 0\\
                                  0 & \sqrt{1-S}   
                                \end{bmatrix}, \qquad \bm{\mathcal{P}}_1 = \begin{bmatrix}
                                                                                                            0 & 0\\
                                                                                                             0 & \sqrt{S}   
                                                                                                \end{bmatrix}.
\end{equation}
The parameter $S$ can be modeled by the following time dependence $S = 1 - \cos^2(\chi t)$, for $0 \le \chi t \le \pi/2$. The quantumness parameter, in this case, is given by
\begin{equation}\label{QPD}
Q(\Phi^{PD}) = 1- S = \cos^2(\chi t).
\end{equation}
The average fidelity turns out to be 
\begin{equation}\label{avgFidPD}
\mathcal{F}_{PD} = \frac{1}{3} [2 + \cos(\chi t)].
\end{equation}
\begin{figure}
	\centering
	\includegraphics[width=60mm]{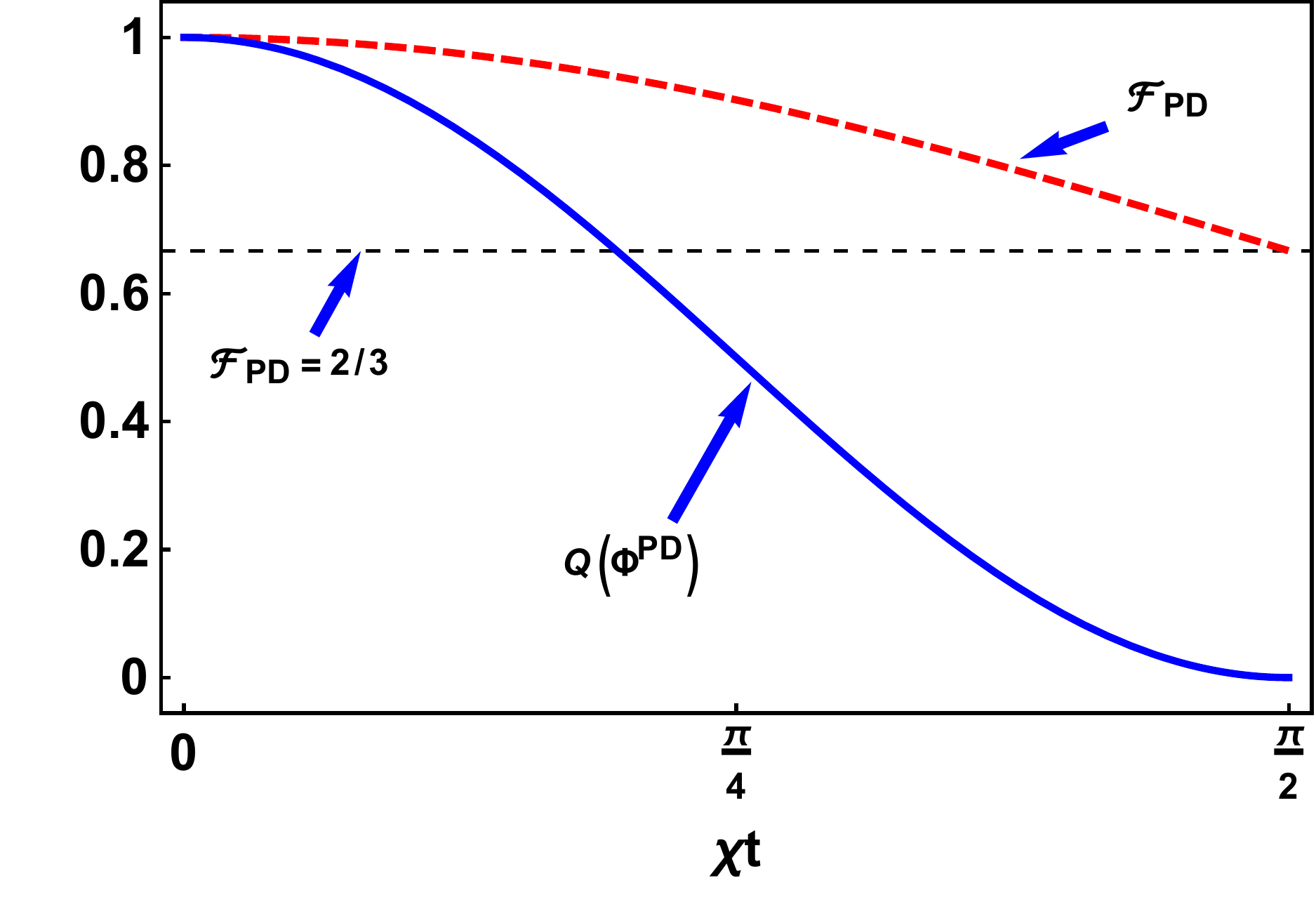}
	\caption{PD channel: The quantumness $Q(\Phi^{PD})$ Eq. (\ref{QPD}) and \textit{average} fidelity $\mathcal{F}_{PD}$ Eq. (\ref{avgFidPD}), plotted with respect to the dimensionless quantity $\chi t$, for a qubit subjected to PD noise.}
	\label{fig:PD}
\end{figure}

\textit{Generalized Amplitude Damping (GAD) channel}: GAD is a generalization of the AD channel to finite temperatures \cite{banerjee2018open}. The later models  processes like spontaneous emission from an atom and is pertinent to the problem of quantum erasure \cite{srikanth2007environment}. The dynamics, in this case, is governed by the following Kraus operators
\small
\begin{align}\label{eq:GAD}
\bm{\mathcal{A}}_0 &= \begin{bmatrix}
                                    \sqrt{\Theta} &  0\\
                                    0                 & \sqrt{s \Theta}
                                 \end{bmatrix},   \quad           \bm{\mathcal{A}}_1 = \begin{bmatrix}
                                                                                                                     0      &  \sqrt{p \Theta}\\
                                                                                                                     0      &  0
                                                                                                        \end{bmatrix},       \nonumber \\  
\bm{\mathcal{A}}_2 &= \begin{bmatrix}
                              \sqrt{s (1-\Theta)} &  0\\
                                       0        & \sqrt{ (1-\Theta)}
                              \end{bmatrix},   \quad           \bm{\mathcal{A}}_3 = \begin{bmatrix}
                                                                                0      &  0\\
                                                                   \sqrt{p (1-\Theta)}      &  0
                                                                             \end{bmatrix}.                                                                                                               
\end{align}
\normalsize
Here, $\Theta = \frac{n + 1}{2 n + 1}$, and $s = \exp[-\frac{\gamma t}{2} (2 n + 1)]$. Also, $n$ is the mean number of excitations in the bath and $\gamma$ represents the spontaneous emission rate. In the zero temperature limit, $n=0$, implying $\Theta = 1$, thereby recovering the AD channel. The quantumness parameter for GAD channel comes out to be 
\begin{equation}\label{QGAD}
Q(\Phi^{GAD}) =  \begin{cases}
  \frac{1}{2}s+\tilde{s} &  {\rm for}~~ t \le \tau,\\\\

s & {\rm for}~~  t > \tau.
\end{cases} 
\end{equation}
with,
\begin{align*}
\tilde{s} &= \frac{5}{2}  ( 2\Theta - 1 )^2 (1-s)^2,\nonumber\\
\tau    &= - \frac{2}{\gamma (2 n + 1)} \ln \bigg[ \frac{5}{6 + 4 n + n^2} \bigg].\\
\end{align*}

\begin{figure}[ht]
	\centering
	\includegraphics[width=60mm]{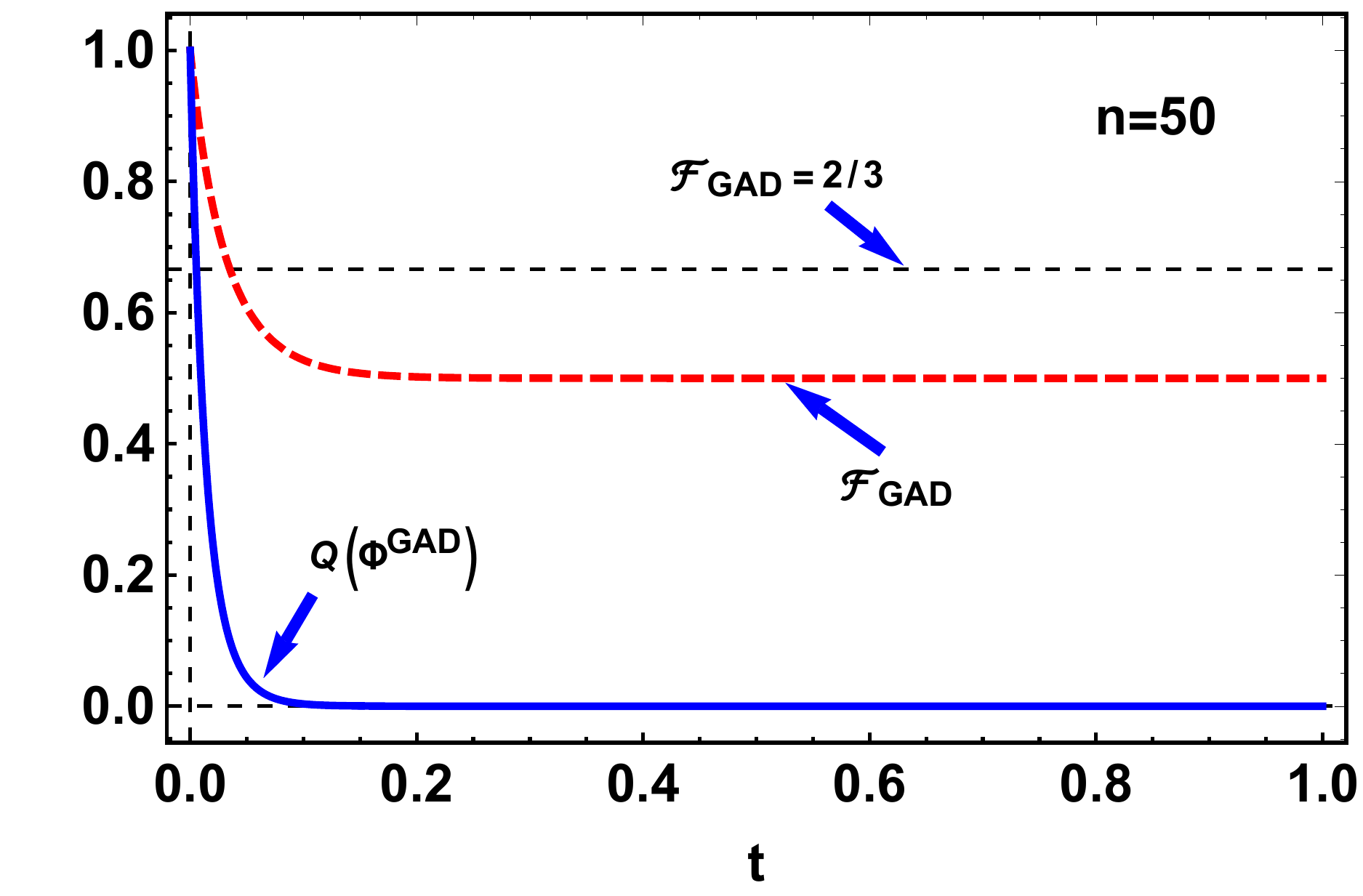}\\
	\includegraphics[width=60mm]{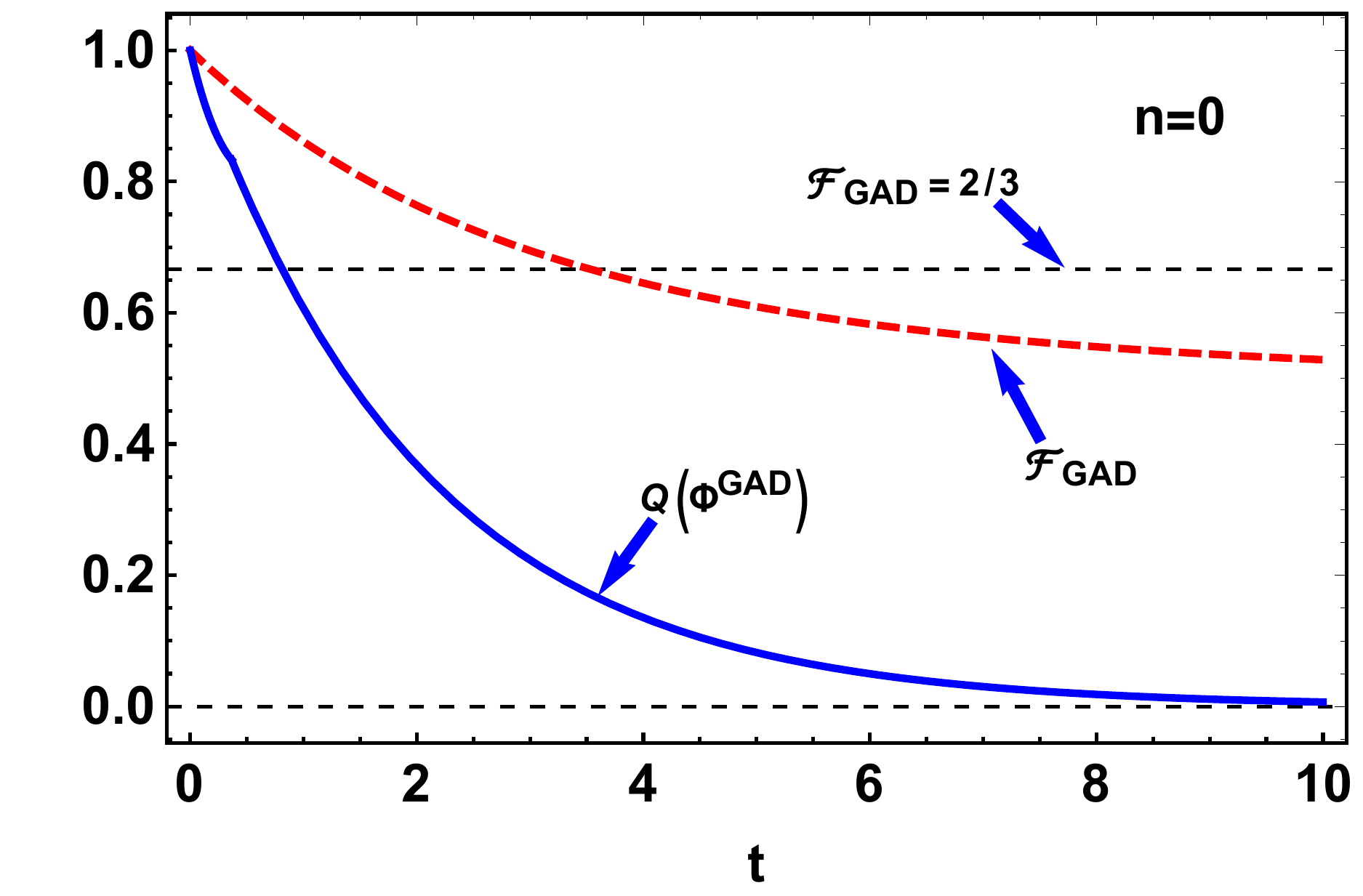}
	\caption{GAD channel: The quantumness $Q(\Phi^{GAD})$ Eq. (\ref{QGAD}) and fidelity $\mathcal{F}_{GAD}$ Eq. (\ref{avgFidGAD}), plotted with respect to time $t$ (sec.), for a qubit subjected to GAD noise. With $\gamma =1$, the top and bottom panels correspond to the cases when $n=50$ and $0$, respectively. Here,  $\tau \approx 0.1246 $ and $ 0.3646$ in the former and later case, respectively . The $n=0$ case corresponds to the zero temperature limit, such that GAD reduces to AD noise.}
	\label{fig:GAD}
\end{figure}
The average fidelity in this case is given by
\small 
\begin{equation}\label{avgFidGAD}
\mathcal{F}_{GAD} = \frac{1}{6} [3 + 2 \sqrt{s} + s].
\end{equation}
Here $s = \exp[-\frac{\gamma t}{2} (2 n + 1)]$.
\normalsize

\par
\textit{Unruh channel:} To an  observer undergoing acceleration $a$, the Minkowski vacuum appears as a warm gas emitting black-body radiation at temperature given by $T = \frac{\hbar a}{2 \pi c K_B}$, called the  \textit{Unruh temperature} and the effect is known as the \textit{Unruh effect}. The Unruh effect has been described as a noisy quantum channel with the following Kraus operators
\begin{equation}
\mathbf{U}_0 = \begin{bmatrix}
                            \cos(r)  &   0\\
                            0          &  1        
                       \end{bmatrix}  \qquad {\rm and } \qquad   \mathbf{U}_1 = \begin{bmatrix}
                                                                                                                        0  &   0\\
                                                                                                                \sin(r)   &  0        
                                                                                                               \end{bmatrix}.
\end{equation} 
Here, $ \cos(r) = [ 1 + \exp(-2\pi c \omega / {\rm a})]^{-1/2}$. The quantumness parameter for the Unruh channel turns out to be
\begin{equation}\label{Qunruh}  
Q(\Phi^{{\rm Unruh}}) = \cos^2(r).
\end{equation}
The average fidelity here is
\begin{align}\label{avgFidunruh}
\mathcal{F}_{{\rm Unruh}} &= \frac{1}{12} (4 \cos (r)+\cos (2 r)+7).
\end{align}

\begin{figure}[ht!]
	\centering
	\includegraphics[width=60mm]{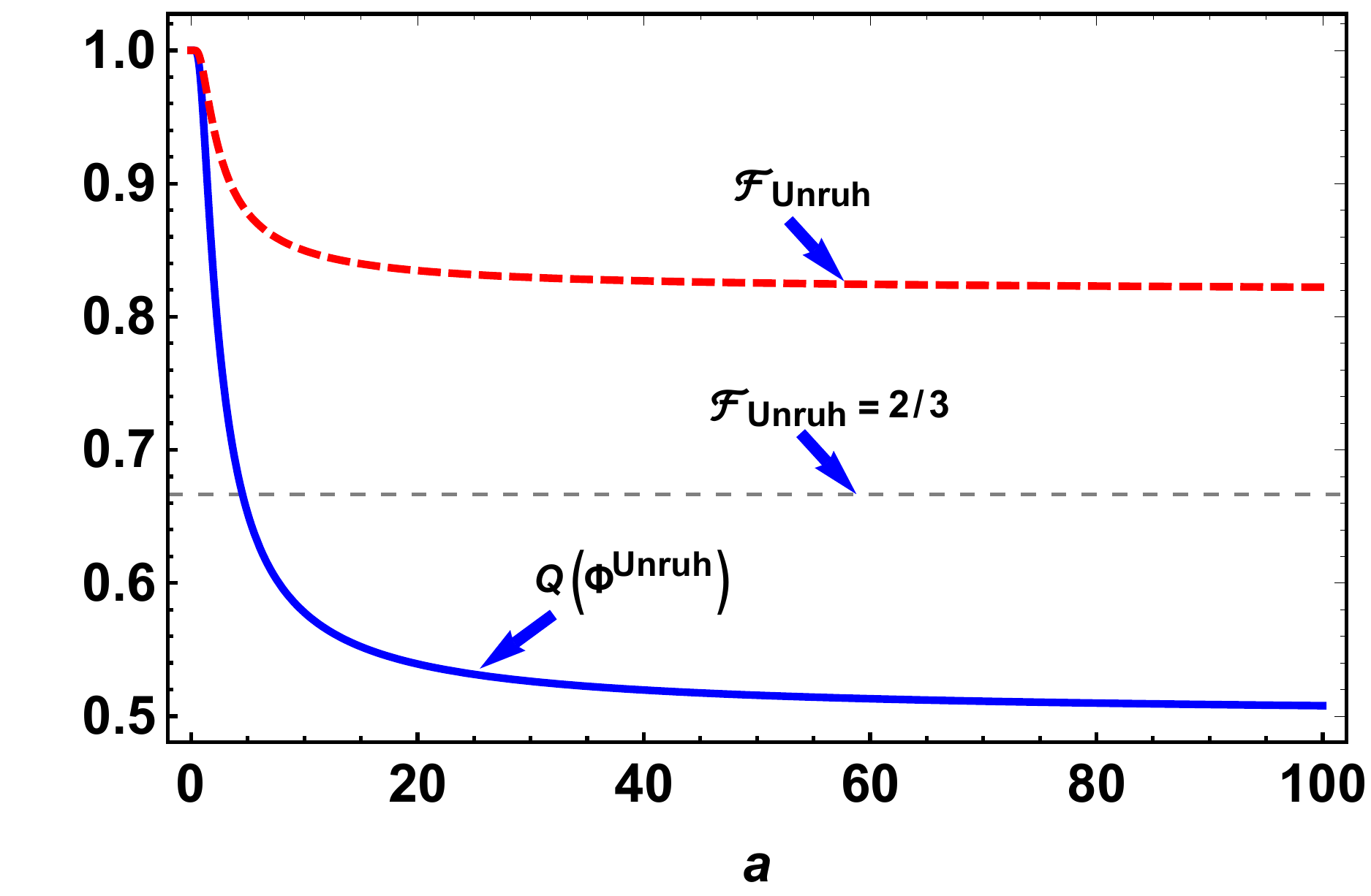}\\
	\caption{Unruh channel: The behavior of quantumness and average fidelity depicted with respect to the acceleration a (in  units $\hbar = c = 1$).}
	\label{fig:Unruh}
\end{figure}

\section{Results and discussion}\label{resultdiscussion}
The quantumness of noisy channels is quantified by the coherence measure given by Eq. (\ref{generalQ}). For specific case of a two level system (qubit), using square $l_1$ norm as a measure of coherence, one obtains a simple working rule for computing the quantumness of a channel, given in Eq. (\ref{Qdef}).\par
For RTN channel, the quantumness measure turns out to be the square of  the memory kernel $\Lambda (t)$, defined in Eq. (\ref{Lambdadef}). In the non-Markovian regime, both the quantumness as well as fidelity are seen to sustain much longer in time as compared with the Markovian case, Fig. \ref{fig:RTN}. In the limit $t \rightarrow \infty$, $\Lambda (t) \rightarrow 0$, consequently, we have
\begin{equation}
Q(\Phi^{RTN}) =  \Lambda^2(t) \rightarrow 0,~~ {\rm and}~~ \mathcal{F}_{RTN} = \frac{1}{3} (2 + \Lambda(t)) \rightarrow \frac{2}{3}.
\end{equation}
This is consistent with our notion of fidelity less than or equal to $2/3$ for a processes that can be simulated by a classical theory.
The NMD channel shows non-zero quantumness within the allowed range, i.e., $[0, 1/2]$, of time like parameter $p$, for $0< \alpha \le 1 $. In this case, the parameter $\alpha$ quantifies the degree of non-Markovianity, which increases as $\alpha$ goes from $0$ to $1$. At $p=1/2$, i.e., for $t \rightarrow \infty$, $\Omega (p) =- \alpha/2$, we have
\begin{equation}
Q(\Phi^{NMD}) = \alpha^2/4 \quad {\rm and}\quad \mathcal{F}_{NMD} = \frac{2}{3} (1 -  \alpha/2)
\end{equation}
That is, the quantumness parameter is always positive but the average fidelity goes below its classical limit. This is consistent with \cite{Shahbeigi2018} that a nonzero value of the coherence based measure of quantumness is a necessary but not sufficient criterion for quantum advantage in teleportation fidelity.  However, in the Markovian limit, i.e., $\alpha \rightarrow 0$, $\Omega \rightarrow 1-2p$. Since $p \in [0,1/2]$, implies $\Omega \in [0,1]$. Therefore, $Q(\Phi^{NMD}) = (1-2p)^2$ and $ \mathcal{F}_{NMD} = \frac{1}{3} [2 + (1-2p)]$,  both the quantities lead to similar predictions in this limit. These features are depicted in Fig. \ref{fig:NMD}. 

 One of the purely quantum noise channels is the PD channel which characterizes the processes accompanied with the loss of coherence without loss of energy. The behavior of quantumness and average fidelity, in this case, is depicted in Fig \ref{fig:PD}. The parameter $Q(\Phi^{PD})$ becomes zero as  $\mathcal{F}_{PD}$ reaches $2/3$.\par
Next we analyzed   non-dephasing models such as GAD and Unruh channels. From the GAD channel, one can recover the AD channel in the zero temperature limit, i.e., when $n = 0$, see Eq. (\ref{eq:GAD}). In this case $\Theta = 1$ and the quantumness parameter, with $\xi = 1-s$, becomes
\begin{equation}\label{QforAD}
Q(\Phi^{AD}) =  \begin{cases}
\frac{1}{2} [6 \xi^2 - 3 \xi +2] &  {\rm for}~~ \xi \le 1/6,\\\\

1- \xi & {\rm for}~~  \xi > 1/6.
\end{cases} 
\end{equation}
This is consistent with the results given in \cite{Shahbeigi2018}.
 In the case of GAD channel, the quantumness parameter is nonzero even though the average fidelity goes below its classical limit $2/3$. This reiterates the statement made earlier regarding quantumness and average teleportation fidelity, Fig. \ref{fig:GAD}. In the high temperature regime, both the measures, i.e., quantumness as well as average fidelity seem to lead to  similar predictions at the same time. For Unruh channel, the quantumness and average fidelity are studied with respect to the acceleration $a$. Both the measures show a saturation at values which are well above their classical limits, Fig \ref{fig:Unruh}. This is in consonance with \cite{UnruhSB}, where it was shown that the Unruh channel, though structurally similar to the AD channel, is different from it.

\section{Conclusion}\label{conclusion}
We studied the quantumness and average fidelity of various channels, both Markovian as well as non-Markovian. Specifically, we considered the dephasing channels like RTN, NMD and PD channels and non-dephasing channels such as GAD and Unruh channels. The non-Markovian dynamics (exhibited by RTN and NMD channels in this case)  is found to favor the nonclassicality. This is explicitly seen from the fact that a nonzero value of  parameters controlling the degree of non-Markovianity takes the quantumness  beyond the classical value. The non-Markovianity assisted enhancement of nonclassicality can be of profound importance in carrying out  quantum information tasks. This can be realized by effectively  engineering  the  system-reservoir models.   The quantumness measure and average fidelity exhibit  similar predictions for the Unruh channel. Similar behavior is observed for the dephasing channels, albeit, in the  Markovian regime. This can bee seen in RTN and NMD channels. In contrast, in the non-dephasing  Lindbladian channel, considered here, the quantumness witness  and average fidelity show qualitatively similar results.  
\par
Such a study of the interplay between nonclassicality of the quantum channels with the underlying dynamics can be useful from the quantum information point of view,  and also brings out the effectiveness of the measure of quantumness under different types of dynamics.

\section*{Acknowledgment}
We thank Prof. R. Srikanth of PPISR, Bangaluru, India, for useful discussions during the preparation of this manuscript. 

\end{document}